\documentclass[final,1p,times]{elsarticle} 

\usepackage{graphicx}
\usepackage{amssymb} 
\usepackage{amsthm} 
\usepackage{lineno}

\journal{Nuclear Physics A} 

\begin{document}

\begin{frontmatter} 

\title{In-Medium Quarkonia at SPS, RHIC and LHC}

\author[auth1]{X.~Zhao}
\author[auth2]{A.~Emerick}
\author[auth3]{R.~Rapp}
\address[auth1]{Department of Physics and Astronomy, Iowa State University,
Ames, IA 50011, USA}
\address[auth2]{School of Physics and Astronomy, University of Minnesota, Minneapolis, MN 55414, USA}
\address[auth3]{Cyclotron Institute and Dept. of Physics \& Astronomy,
Texas A\&M University, College Station, TX 77843-3366, USA}

\begin{abstract} 
A kinetic-rate equation approach in a thermally expanding medium is employed to 
calculate the evolution of charmonium and bottomonium distributions in heavy-ion 
collisions. The equilibrium properties of the quarkonia are taken from in-medium
spectral functions which are schematically constrained by euclidean correlators 
from lattice QCD. The initial conditions for the rate equation (heavy-flavor cross 
sections, nuclear absorption) and the thermal evolution are constrained by 
data as available. After fixing two free parameters to describe 
charmonium data at SPS 
and RHIC, the predictions for LHC are discussed in light of recent data.  
\end{abstract} 

\end{frontmatter} 


\section{Introduction}
Heavy quarkonia are an excellent tool to study the modifications of the 
basic QCD force in hot/dense matter. In vacuum, potential models provide 
a theoretically controlled and phenomenologically successful spectroscopy, 
serving as a calibrated starting point to utilize quarkonium properties 
in medium as a 
probe~\cite{Rapp:2008tf,Kluberg:2009wc,BraunMunzinger:2009ih,Bazavov:2009us}.
However, the realization of such a program in ultra-relativistic
heavy-ion collisions (URHICs) is a non-trivial task. The interplay of
color screening and dissociation reactions in a locally equilibrated, yet
evolving medium, as well as off-equilibrium aspects in initial conditions
and in heavy-quark (HQ) kinetics, result 
in a rather complex problem. A viable theoretical framework to properly 
interpret experimental data must therefore be able to account for these aspects, 
while maintaining the connection to theoretically calculated in-medium properties. 
In particular, it should be able to encompass both charmonium and
bottomonium observables, and a large range in collision energies, as are 
now available. 

In Sec.~\ref{sec_rate-eq} we recall our kinetic-rate equation approach to 
quarkonia in URHICs~\cite{Grandchamp:2003uw,Zhao:2010nk}, its inputs and 
parameters. In Sec.~\ref{sec_pheno} we discuss applications to charmonium 
(\ref{ssec_psi}) and bottomonium (\ref{ssec_ups}) data, mostly comparing 
predictions to recent LHC data. A short summary is given in Sec.~\ref{sec_sum}.

\section{Kinetic Rate Equation}
\label{sec_rate-eq}
Starting from the Boltzmann equation 
one can integrate out the space and 3-momentum dependencies to obtain 
a rate equation for the quarkonium number, $N_{\cal{Q}}$, 
\begin{equation}
\frac{dN_{\cal{Q}}}{d\tau} = -\Gamma_{\cal{Q}} \  
( N_{\cal{Q}} - N_{\cal{Q}}^{\rm eq} ) \ , 
\end{equation}
where ${\cal{Q}}=\psi,\Upsilon$.
This form renders the relation to the ``transport" parameters particularly 
transparent. The equilibrium abundance, $N_{\cal{Q}}^{\rm eq}$, depends on the 
quarkonium mass, $m_{\cal{Q}}$ and binding energy, $\varepsilon_B^{\cal{Q}}$, 
as well as on the HQ mass, $m_Q$, via relative chemical equilibrium at fixed 
HQ number.  The temperature dependence of these quantities is taken from a 
$T$-matrix approach~\cite{Riek:2010fk}. In addition, $N_{\cal{Q}}^{\rm eq}$ 
is corrected for off-equilibrium HQ distributions (which suppress the formation 
rate) using a relaxation-time approximation (which has been verified to work 
well~\cite{Song:2012at}). The pertinent $\tau_Q^{\rm eq}$ is one of the parameters 
of the approach. The type of the inelastic reaction rate, $\Gamma_{\cal{Q}}$,
depends on the relation of $\varepsilon_B^{\cal{Q}}$ to $T$. For 
$\varepsilon_B^{\cal{Q}}<T$, quasifree dissociation prevails (Landau damping of 
the exchanged gluon), while for larger $T$ gluo-dissociation
(singlet-to-octet transitions) takes over. We  adopt the former (latter)
for charmonia (bottomonia), where the precise value of the coupling constant
is utilized as a second fit parameter.
The resulting charmonium~\cite{Zhao:2010nk} and bottomonium~\cite{Emerick:2011xu} 
spectral functions are checked against euclidean correlators from lattice QCD. 
The initial conditions of the rate equation consist of the numbers of 
(would-be formed) quarkonia and heavy quarks, corrected for ``cold-nuclear-matter" 
effects (shadowing, Cronin effect and nuclear absorption).
They are determined from available experimental cross sections or interpolations 
thereof. The evolving medium is modeled by an expanding thermal fireball 
constrained by hadro-chemistry, hadron yields 
and spectra. The quarkonium ground and excited states are evolved with their
individual binding energies and reaction rates, to account for feeddown.    

\section{Quarkonium Phenomenology in URHICs}
\label{sec_pheno}
The simultaneous study of charmonium and bottomonium observables in URHICs 
is particularly valuable, since the differences in charmonium and bottomonium
binding (aka screening effects), as well as in open-charm and open-bottom 
content (aka reaction rates) enable a much improved discrimination power of 
bound-state suppression and regeneration mechanisms. 

\subsection{Charmonium}
\label{ssec_psi}
In Ref.~\cite{Zhao:2010nk} we have employed the above-described framework to 
conduct a systematic analysis of $J/\psi$ data in Pb-Pb and Au-Au collisions 
at top SPS ($\sqrt{s}$=17.3\,AGeV) and RHIC ($\sqrt{s}$=200 AGeV) energies, 
respectively. We have 
investigated both a strong- and a weak-binding scenario, where the former 
describes the data slightly better and gives values for our two parameters 
consistent with the $T$-matrix approach~\cite{Riek:2010fk}, i.e., 
$\alpha_s$$\simeq$0.3 in the quasifree dissociation rate and 
$\tau_c^{\rm eq}$$\simeq$5\,fm/$c$ for the kinetic charm-quark relaxation time. 
We will therefore focus on this scenario from here on. Predictions have then 
been made for LHC and lower RHIC energies. For the latter, the centrality 
dependence of the total nuclear modification factor, $R_{AA}(N_{\rm part})$,
has little $\sqrt{s}$ dependence, consistent with recent PHENIX 
data~\cite{Adare:2012wf}.
However, the composition of the $J/\psi$ gradually changes with an increase
in both the suppression of the primordial component and the regeneration,
reaching ca.~30\% of the total in central Au-Au.  

\begin{figure}[htbp]
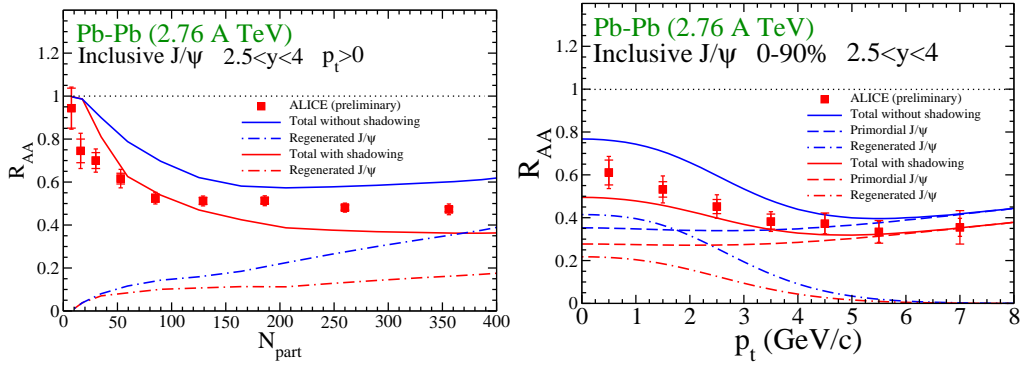

\includegraphics[width=0.49\textwidth]{raa-npart.eps}
\includegraphics[width=0.49\textwidth]{raa-pt.eps}
\vspace{-0.2cm}
\caption{Predictions of the kinetic-rate equation approach~\cite{Zhao:2011cv} 
for the inclusive $J/\psi$ nuclear modification factor in Pb-Pb(2.76\,ATeV) as 
a function of centrality (left) and transverse momentum (right), compared to 
ALICE data~\cite{Suire:2012gt}. Blue and red lines correspond to a charm cross 
section (per unit rapidity) of 0.5\,mb and 0.33\,mb, respectively.}
\label{fig_jpsi}
\end{figure}
The ``degeneracy" in the SPS/RHIC regime was predicted to be broken at the 
LHC~\cite{Zhao:2011cv}, due to the increase in charm cross section which
enhances regeneration more than primordial suppression.  
The agreement with recent ALICE dimuon data (at somewhat forward rapidity)
is fair, cf.~Fig.~\ref{fig_jpsi}. In particular, the tell-tale signature of
the regeneration component at low transverse momentum is confirmed.  
Very similar results are obtained in the transport approach of 
Ref.~\cite{Liu:2009nb}, which differs in details of the implementation, 
but overall asserts the robustness of the conclusions. The statistical
hadronization model also accounts for the centrality dependence of the $J/\psi$ 
yield~\cite{Andronic:2006ky}, using somewhat smaller charm cross sections than 
used in the transport models. This is due to the off-equilibrium effects in 
the latter.  

\begin{figure}[htbp]
\begin{minipage}{0.5\textwidth}
\vspace{-0.3cm}
\includegraphics[width=1.15\textwidth]{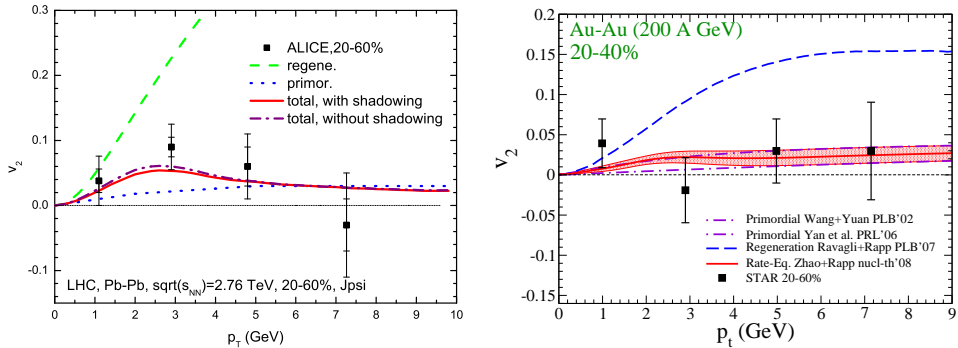}
\end{minipage}
\hspace{0.3cm}
\begin{minipage}{0.5\textwidth}
\includegraphics[width=0.9\textwidth]{v2-jpsi-rhic.eps}
\end{minipage}
\vspace{-0.5cm}
\caption{Elliptic flow of inclusive $J/\psi$ in semicentral Pb-Pb(2.76\,ATeV) 
(left)~\cite{He:2012} and Au-Au(0.2\,ATeV) (right), compared to 
ALICE~\cite{Suire:2012gt} and STAR~\cite{Qiu:2012zz} data, respectively.}
\label{fig_v2}
\end{figure}

Another signature of regenerated charmonia is the collectivity that they inherit
from the charm quarks, which is much larger than from path-length type suppression
effects. However, as is well known, a large radial flow suppresses the $v_2$
of heavy particles at low $p_t$, which is unfortunately where the regeneration
contributes most. Nevertheless, first LHC data show a promising signal
(Fig.~\ref{fig_v2} left) while at RHIC the current data accuracy does not
permit a definite conclusion yet (Fig.~\ref{fig_v2} right).

\subsection{Bottomonium}
\label{ssec_ups}
\begin{figure}[tbp]
\begin{center}
\includegraphics[width=0.48\textwidth]{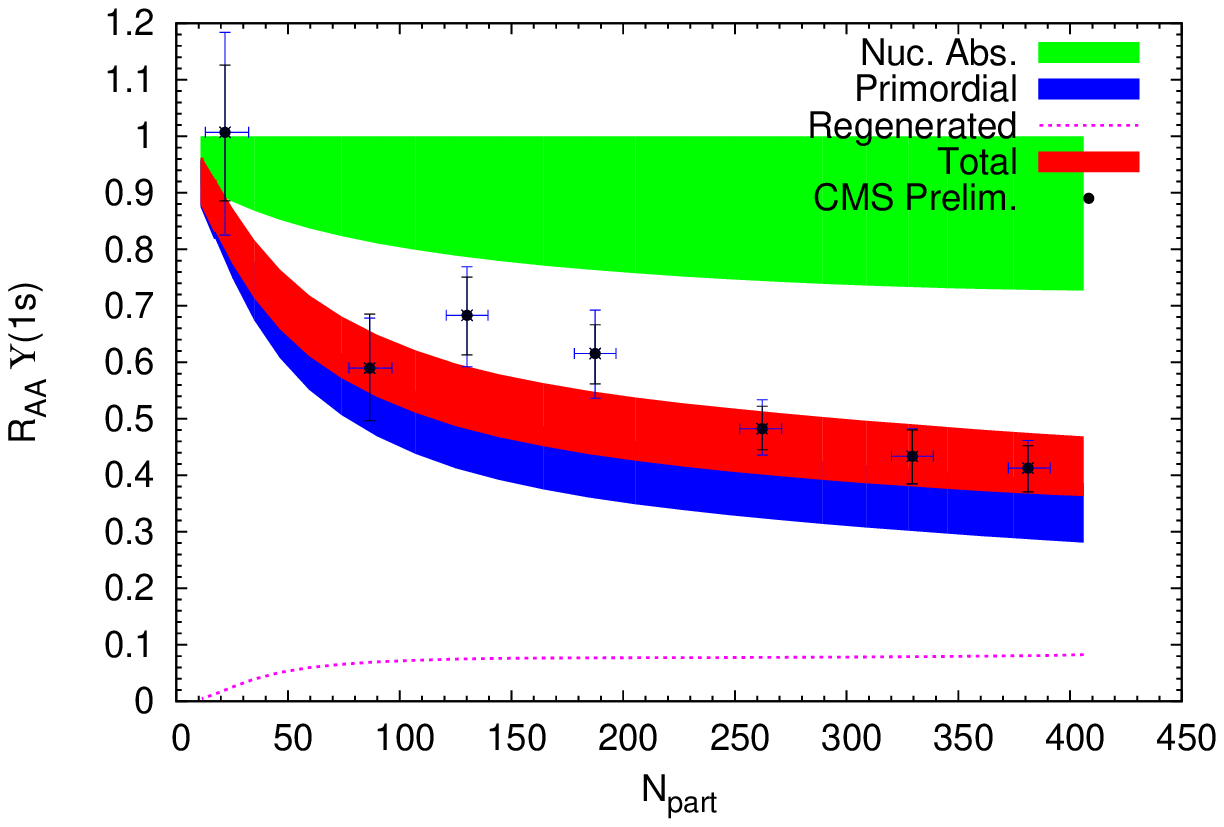}
\includegraphics[width=0.48\textwidth]{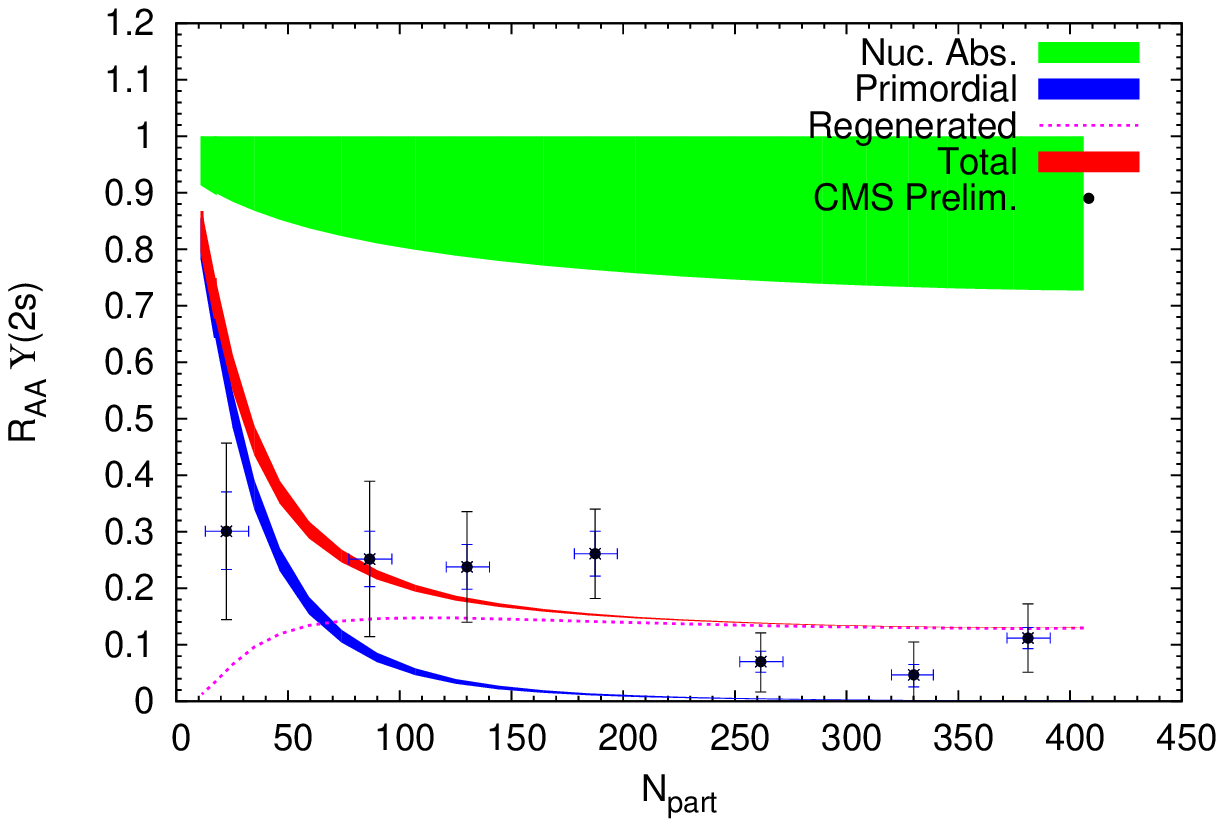}
\end{center}
\vspace{-0.5cm}
\caption{Centrality dependence of nuclear modification factors for inclusive
$\Upsilon(1S)$ (left) and $\Upsilon(2S)$ (right)~\cite{Emerick:2011xu}, 
compared to CMS data~\cite{Chatrchyan:2012fr}.}
\label{fig_ups}
\end{figure}

Utilizing the rate-equation approach for bottomonium production, their degree
of suppression has been identified as a rather sensitive measure of color 
screening, in connection with appropriate dissociation 
reactions~\cite{Grandchamp:2005yw}. This has recently been revisited using 
updated input for open-bottom and bottomonium cross sections at 
2.76\,TeV~\cite{Emerick:2011xu}. Indeed, the resulting $\Upsilon(1S)$ $R_{AA}$ 
agrees reasonably well with CMS data in the strong-binding scenario 
(Fig.~\ref{fig_ups}), but is suppressed too much in the weak-binding scenario. 
On the other hand, $\Upsilon(2S)$ is strongly suppressed even in the former 
(Fig.~\ref{fig_ups} right), with the finally observed yield ascribed to 
regeneration. This should be tested in the $p_t$ spectra. 

\section{Summary}
\label{sec_sum}
A kinetic-rate equation approach, implementing in-medium quarkonium 
properties into a thermal bulk-evolution model, is found to provide a suitable 
tool (with predictive power) to interpret $J/\psi$ and $\Upsilon$ observables
in URHICs. In particular, a gradual increase of regeneration contributions
to $J/\psi$ production from SPS via RHIC to LHC is supported 
by recent LHC data, showing an increase in $R_{AA}$. The associated low-$p_t$ 
enhancement and $v_2$ signal corroborate this interpretation. Bottomonium
observables, while possibly not free of regeneration, give a more direct
access to the in-medium QCD force; the moderate suppression of $\Upsilon(1S)$ 
at LHC (and RHIC) suggests that a strong binding potential persists into
the QGP, which may well be related to a small HQ diffusion coefficient. 
Systematic improvements of the approach are in progress.
\\

{\bf Acknowledgment} Work supported by US-NSF grant no.~PHY-0969394,
NSF-REU grant no.~PHY-1004780, and US-DOE grant no.~DE-FG02-87ER40371.

\end{document}